\documentstyle[12pt]{article}

\def\hybrid{\topmargin 0pt      \oddsidemargin 0pt
        \headheight 0pt \headsep 0pt
        \voffset=-0.5cm
        \textwidth 6.25in       
        \textheight 9.5in       
        \marginparwidth 0.0in
        \parskip 5pt plus 1pt   \jot = 1.5ex}
\catcode`\@=11
\def\marginnote#1{}

\newcount\hour
\newcount\minute
\newtoks\amorpm
\hour=\time\divide\hour by60
\minute=\time{\multiply\hour by60 \global\advance\minute by-\hour}
\edef\standardtime{{\ifnum\hour<12 \global\amorpm={am}%
        \else\global\amorpm={pm}\advance\hour by-12 \fi
        \ifnum\hour=0 \hour=12 \fi
        \number\hour:\ifnum\minute<10 0\fi\number\minute\the\amorpm}}
\edef\militarytime{\number\hour:\ifnum\minute<10 0\fi\number\minute}

\def\draftlabel#1{{\@bsphack\if@filesw {\let\thepage\relax
   \xdef\@gtempa{\write\@auxout{\string
      \newlabel{#1}{{\@currentlabel}{\thepage}}}}}\@gtempa
   \if@nobreak \ifvmode\nobreak\fi\fi\fi\@esphack}
        \gdef\@eqnlabel{#1}}
\def\@eqnlabel{}
\def\@vacuum{}
\def\draftmarginnote#1{\marginpar{\raggedright\scriptsize\tt#1}}

\def\draftlabel#1{{\@bsphack\if@filesw {\let\thepage\relax
   \xdef\@gtempa{\write\@auxout{\string
      \newlabel{#1}{{\@currentlabel}{\thepage}}}}}\@gtempa
   \if@nobreak \ifvmode\nobreak\fi\fi\fi\@esphack}
        \gdef\@eqnlabel{#1}}
\def\@eqnlabel{}
\def\@vacuum{}
\def\draftmarginnote#1{\marginpar{\raggedright\scriptsize\tt#1}}

\def\draft{\oddsidemargin -.5truein
        \def\@oddfoot{\sl preliminary draft \hfil
        \rm\thepage\hfil\sl\today\quad\militarytime}
        \let\@evenfoot\@oddfoot \overfullrule 3pt
        \let\label=\draftlabel
        \let\marginnote=\draftmarginnote
   \def\@eqnnum{(\theequation)\rlap{\kern\marginparsep\tt\@eqnlabel}%
\global\let\@eqnlabel\@vacuum}  }

\def\underline#1{\relax\ifmmode\@@underline#1\else
        $\@@underline{\hbox{#1}}$\relax\fi}

\def\titlepage{\@restonecolfalse\if@twocolumn\@restonecoltrue\onecolumn
     \else \newpage \fi \thispagestyle{empty}\c@page\z@
        \def\thefootnote{\fnsymbol{footnote}} }

\def\endtitlepage{\if@restonecol\twocolumn \else  \fi
        \def\thefootnote{\arabic{footnote}}
        \setcounter{footnote}{0}}  
\relax

\hybrid

\def\beq{\begin{equation}}
\def\eeq{\end{equation}}
\def\p{\partial}

\def\DD{{\sf D}}
\def\Dc{{\sf D^c}}

\begin{document}

\begin{titlepage}

\title{Large $N$ expansion  for
normal and complex matrix ensembles}

\author{P. Wiegmann \thanks{James Frank Institute and Enrico Fermi
Institute
of the University of Chicago, 5640 S.Ellis Avenue,
Chicago, IL 60637, USA and
Landau Institute for Theoretical Physics, Moscow, Russia}
\and A. Zabrodin
\thanks{Institute of Biochemical Physics,
Kosygina str. 4, 119991 Moscow, Russia
and ITEP, Bol. Cheremushkinskaya str. 25, 117259 Moscow, Russia}}

\date{September 2003}
\maketitle

\begin{abstract}

We present the first two leading terms of
the $1/N$ (genus) expansion
of the free energy
for ensembles of normal and complex random matrices.
The results are expressed through the support of eigenvalues
(assumed to be a connected domain
in the complex plane).
In particular,
the subleading (genus-1)  term
is given by the regularized determinant of the Laplace
operator in the complementary domain
with the  Dirichlet boundary conditions.
An explicit expression of the genus
expansion through  harmonic moments of the domain
gives some new representations
of the mathematical objects related to the
Dirichlet boundary problem, conformal analysis and
spectral geometry.

\end{abstract}

\vfill

\end{titlepage}

\section{Introduction}

Ensembles of random matrices
have numerous important
applications in physics and mathematics ranging from
energy levels of nuclei to number theory.
An important information is encoded in the
$1/N$ expansion ($N$ is the size of the matrix)
of different expectation values in the ensemble.
Many relevant references can be found in \cite{FSV} .

In this paper we discuss $1/N$-expansion in
statistical
ensembles of normal and complex matrices.
A matrix $M$ is called normal if it commutes with
its Hermitian conjugate: $[M, \, M^{\dag}]=0$, so both
matrices can be diagonalized simultaneously.
Eigenvalues of a normal matrix are complex
numbers. The statistical weight
$$e^{\frac{1}{\hbar}\,{\rm tr} \, W(M)}d \mu (M)$$
of the normal matrix
ensemble is specified by a potential function
$W(M)$ (which depends on both $M$ and $M^{\dag}$).
Here $\hbar$ is a parameter,
and the measure $d\mu$ of integration over normal matrices
is induced from the flat metric on the space
of all complex matrices.

Along the standard procedure of integration over angle
variables \cite{Mehta},
one passes to the joint probability distribution of eigenvalues
$z_1 , \ldots , z_N$.
The partition function is then
given by the integral
\beq\label{Z1}
Z_N=\frac{1}{(2\pi^3 \hbar )^{N/2} N!}
\int |\Delta_N (z_i)|^2 \prod_{j=1}^{N}
e^{\frac{1}{\hbar}W(z_j )}d^2 z_j
\eeq
Here $\Delta_N (z_i) =\prod_{i>j}^N (z_i -z_j )$
is the Vandermonde determinant and $d^2 z \equiv dx \, dy$
for $z=x+iy$. The $N$-dependent normalization factor
is put here for further convenience.

The model of normal matrices was introduced
in \cite{Zabor}.
This model is the particular $\beta =1$ case of a
more general one, referred to as 2D Coulomb gas
with the joint probability distribution
$|\Delta_N (z_i)|^{2\beta} \prod_{j=1}^{N}
e^{\frac{1}{\hbar}W(z_j )}d^2 z_j$.

For the potential of the form
\beq\label{Z2}
W(z)=-z\bar z +V(z)+\overline{V(z)}
\eeq
where $V(z)$ is an analytic function in some region
of the complex plane (say, a polynomial),
the normal matrix model is equivalent to the ensemble
of all complex matrices with the same potential.
It generalizes the gaussian Ginibre-Girko ensemble \cite{GG}.
When passing to the integral over eigenvalues,
the partition function for complex matrices differs
from the one for normal matrices
by a normalization factor only \cite{Mehta}.
Both models are then reduced to the 2D Coulomb gas
(with $\beta =1$) in the external potential.
Note also a formal similarity with the model
of two Hermitian matrices. Its partition function is given
by the same formula (\ref{Z1}), with the potential (\ref{Z2}),
but $z_i$ and $\bar z_i$ are to be regarded as
two independent {\it real} integration variables, with
$d^2 z_i$ being understood as $dz_i d\bar z_i$.

It appears that the potential of the form (\ref{Z2})
is most important for applications \cite{WZ2}.
In the main part of the paper, we concentrate on this case,
so one may, in this context, ignore
the difference between the normal and complex ensembles,
taking the 2D Coulomb gas partition function as a
starting point.
Physical applications of this
model include the quantum Hall effect, the
Saffman-Taylor viscous fingering and, conjecturally,
more general growth problems which are mathematically
described as a random evolution in the moduli space
of complex curves. Recently, the normal matrix model
was shown \cite{AKK} to be closely related to the matrix quantum
mechanics, and, therefore, to the $c=1$ string theory.

In addition to this it appears that the large $N$
limit of the  normal
or complex random matrices admits a natural geometric
interpretation
relevant to the 2D inverse potential problem,
the Dirichlet boundary problem and to spectral geometry
of planar domains.
In this paper we concentrate on calculation of the $1/N$ expansion
of the free energy, $F  \propto \log Z_N$, and on its
algebro-geometric meaning, leaving physical
aspects for future publications.

The large $N$ limit also implies the
limit $\hbar\to 0$, so that
$\hbar N$ is finite and fixed. We prefer to work with
the equivalent $\hbar$-expansion, rather than with
the $1/N$ expansion,
thus emphasizing its semiclassical nature.
The free energy of the Hermitian, two-Hermitian, normal
and complex matrix ensembles with
the potential (\ref{Z2}) has an $\hbar$-expansion
of the form $\log Z_N =\sum_{g\geq 0}\hbar^{2g-2}F_g$, where
$g$-th term is associated with the contribution
of diagrams with Euler characteristics $2-2g$,
in the perturbative expansion of the free energy.
Here we discuss the first two terms, $F_0$ and $F_1$:
\beq\label{Z3}
F= \hbar^2 \log Z_N = F_0 +\hbar^2 F_1 + O(\hbar^4 )
\eeq
The leading term, $F_0$, is the contribution of planar
diagrams, and $F_1$ is commonly refered to as genus 1 correction.

When $N$ becomes large
new macroscopic structures emerge.
Invoking a physical analogy, one may say
that the gas of eigenvalues segregates into ``phases''
with zero and non-zero density separated by a very
narrow interface. The domain $\DD$ in the complex plane
where the density is non-zero is called the support
of eigenvalues (it may consist of several disconnected domains).
The density at any point outside it is exponentially
small as $N\to \infty$.

The leading contribution to the free energy, the $F_0$
term in (\ref{Z3}), is basically the
Coulomb energy of particles confined in the domain $\DD$.
For the potential of the form (\ref{Z2})
it is the tau-function of curves introduced
in \cite{KKMWZ}. It encodes solutions to archetypal
problems of complex analysis and potential theory
in planar domains.

Here we review these results and also compute
the genus-1 correction to the free energy.
The latter is identified with
the free energy of a free bosonic field in the domain $\Dc$
which is complementary to the support
of eigenvalues, i.e., in the domain where the mean density
vanishes:
\beq\label{Z4}
F_1 =-\frac{1}{2}\log \det (-\Delta_{\Dc})
\eeq
Here $\det (-\Delta_{\Dc})$
is a properly regularized determinant of
the Laplace operator in $\Dc$ with Dirichlet
boundary conditions.
This suggests interesting links to
spectral geometry of planar domains.

The genus expansion in the Hermitian random matrix
model beyond
the leading order
has been obtained in the seminal paper \cite{herm1}.
In \cite{Kostov1}, the genus-1 correction was interpreted
in terms of bosonic field theory on a hyperelliptic
Riemann surface.
The genus 1 correction to free energy
of the model of two Hermitian matrices
with polynomial potential was found only recently \cite{Eynard1}.

\section{The planar large $N$ limit}

In this section we briefly recall the large $N$ limit
technique.
This material is standard since early days
of random matrix models (see., e.g., \cite{BIPZ}).
An appealing feature of the
model of normal or complex matrices is a nice geometric
interpretation and a direct relation to the inverse
potential problem in two dimensions.

As was already mentioned, the parameter $\hbar$
tends to zero simultaneously with $N \to \infty$
in such a way that $t_0 =N\hbar $ is kept finite and fixed.
Using the Coulomb gas analogy, one may say that
the leading contribution to the free energy is
equal to the extremal value of the energy
\beq\label{energy}
{\cal E}= \sum_{i\neq j}\log |z_i -z_j | +
\frac{1}{\hbar}\sum_i W(z_i )
\eeq
Equilibrium positions of charges are given by
the extremum of the plasma energy:
$\p_{z_i}{\cal E}=\p_{\bar z_i}{\cal E}= 0$.

Consider the 2D Coulomb potential
$\Phi (z)=-\hbar \sum_i \log |z-z_i|^2$ created by
the charges. Writing it as
$$
\Phi (z)=-\int \log |z-\zeta |^2 \rho (z) \, d^2 z
$$
where
\beq\label{density}
\rho (z)=-\frac{1}{4\pi}\Delta \Phi (z)=
\hbar \sum_i \delta^{(2)}(z-z_i)
\eeq
is the microscopic density of eigenvalues (a sum of two-dimensional
delta-functions), we assume that $\Phi$
in the limit
can be treated as a continuous function.
It is normalized as $\int \Delta \Phi (z)d^2z=-4\pi t_0$.
Let $\Phi_0$ be this function for the equilibrium
configuration of charges, then
\beq\label{semitau1}
\p_z (\Phi_0 (z)-W(z))=
\p_{\bar z} (\Phi_0 (z)-W(z))=0
\eeq
with the understanding that
this equation holds only for
$z$ belonging to a domain (or domains) where the density
is nonzero.
Applying $\p_{\bar z}$ to the both sides, we see that
the equilibrium density, $\rho_0 (z)$,
is equal to $-\frac{1}{4\pi}\Delta W(z)$
in some domain $\DD$
(the support of eigenvalues) and zero
otherwise:
$$
\rho_0 (z)=\left \{
\begin{array}{ll}
 \sigma\;\; & z\in {\sf D}
\\
0\;\; & z\in {\sf D^c}
\end{array}\right.
\;\;\; \mbox{and}  \;\;\;
\Phi_0 (z)=-\int_{\DD}\log |z-\zeta |^2 \sigma\, d^2 \zeta
$$
Here $\Dc ={\bf C}\setminus \DD$ is the domain complimentary
to the support of eigenvalues and
\beq\label{sigma}
\sigma=-\frac{1}{4\pi}\Delta W(z, \bar z)
\eeq
In this and in the next section we
assume the special form of the potential (\ref{Z2}).
Then $\rho_0=1/\pi$ in the domain $\DD$.

The shape of $\DD$ is determined by the function $V(z)$.
Let us assume, without loss of generality, that $0\in \DD$ and
parametrize $V(z)$ by Taylor coefficients at the origin:
\beq\label{V}
V(z)=\sum_{k\geq 1} t_k z^k
\eeq
The parameters $t_k$
(coupling constants of the matrix model)
are in general complex numbers.
Multiplying (\ref{semitau1}) by $z^{-k}$
and integrating over the boundary of $\DD$,
we conclude that
the domain $\DD$ is such that
$-\pi k t_k$'s are moments of its complement, $\Dc$,
with respect to the functions $z^{-k}$:
\beq\label{tk}
t_k =-\, \frac{1}{\pi k}\int_{\Dc}z^{-k}d^2z
=\frac{1}{2\pi ik}\oint_{\p \DD}z^{-k}\bar z\, dz
\eeq
Besides, from the normalization
condition we know that the area of $\DD$ is equal to $\pi t_0$.
To find the
shape of the domain from its moments and area is the subject
of the inverse potential problem. These data
determine it uniquely, at least locally.

Here we assume that $\DD$ is a connected domain.
For example, in the potential $W=-z\bar z$ the eigenvalues
uniformly fill the disk of radius $\sqrt{\hbar N}$.
Small perturbations of the potential slightly disturb
the circular shape.

In what follows, we need some functions associated with
the domain $\DD$, or rather with its complement, $\Dc$.
The basic one is a univalent conformal map from the exterior
of the unit disk onto the domain $\Dc$. Such a map exists
by virtue of the Riemann mapping theorem.
Let ${\sf U}$ be the unit disk and ${\sf U^c}$ its complement,
i.e., the exterior of the unit disk. Consider the
conformal map $z(w)$ from ${\sf U^c}$ onto $\Dc$ normalized
so that $z(\infty )=\infty$ and $r=\lim_{w\to \infty}z(w)/w$ is real,
then the map is unique.
In general, the Laurent expansion of the function
$z(w)$ around infinity is
\beq\label{zw}
z(w)=rw+\sum_{k\geq 0}u_k w^{-k}
\eeq
The real number $r$ is called the (external) conformal radius
of $\DD$. Since the map is conformal, all zeros and poles
of the derivative $z'(w)\equiv \p_w z(w)$ are inside the
unit circle. We also need the function $\bar z (w)$
given by the Laurent series
(\ref{zw}) with complex conjugate coefficients and
the Green function of the Dirichlet boundary problem
in $\Dc$. In terms of the conformal map, the latter
is given by the explicit formula
\beq\label{Green}
G(z, z')=\log \left |
\frac{w(z)-w(z')}{w(z)\overline{w(z')}-1}\right |
\eeq
Here $w(z)$ is the conformal map
from $\Dc$ onto ${\sf U^c}$ inverse to the $z(w)$.

\section{The leading term of the free energy}

The leading contribution to the free energy is the value
of the Coulomb energy (\ref{energy})
(multiplied by $\hbar^2$)
for the extremal configuration of charges:
$$
F_0 =
\int_{\DD}\!
\int_{\DD}\log \left | z -z' \right |
\sigma(z)\sigma(z')d^2 z \,
d^2 z' +\int_{\DD} W(z, \bar z)\sigma d^2 z
$$
The integrated version of the extremum condition
(\ref{semitau1}) tells us that $\Phi_0 (z)-W(z)=
\mbox{const}$ for any $z\in \DD$. The constant
can be found from the same equality at $z=0$, and
we obtain
$F_0$ as an explicit functional of the
domain $\DD$:
\beq\label{F0}
F_0 =-
\int_{\DD}\!
\int_{\DD}\log \left | \frac{1}{z}-\frac{1}{z'}\right |
\sigma(z)\sigma(z')d^2 z \, d^2 z'
\eeq
For the special potential of the form (\ref{Z2}),
when $\sigma =1$, the free energy is
to be regarded as a function of $t_0$
and the coupling constants $t_k$.

Properties of $F_0$
immediately follow from known
correlation functions of the model in the planar
large $N$ limit.
See \cite{WZ2,KKMWZ} for
normal and complex matrices and \cite{Bertola} for similar
results in the context of the Hermitian 2-matrix model.
Some of these correlation functions
previously appeared
in studies of thermal fluctuations in classical confined
Coulomb plasma \cite{Jancovici}.
Integrable structures associated with $F_0$
were studied in \cite{MWWZ,KKMWZ,MWZ,BMRWZ}.
Here is the list of main properties of $F_0$ for
the most important case $\sigma =1/\pi$.

\begin{itemize}
\item
{\it 1-st order derivatives:}
\beq\label{1st}
\begin{array}{l}
\displaystyle{
\frac{\p F_0}{\p t_k}=
\frac{1}{\pi}\int_{\DD} z^k \, d^2 z \,,
\;\;\;\; k\geq 1,}
\\ \\
\displaystyle{
\frac{\p F_0}{\p t_0}=
\frac{1}{\pi}\int_{\DD} \log |z|^2 \, d^2 z }
\end{array}
\eeq
can be combined in the generating formula
\beq\label{1st1}
{\cal D} (z)F_0 =\frac{1}{\pi}\int_{\DD} \log
|z^{-1}-\zeta^{-1} |^2 d^2 \zeta \,,
\;\;\;\; z\in \Dc\,,
\eeq
where
\beq\label{nabla}
{\cal D} (z)=\frac{\p }{\p t_0}+
\sum_{k\geq 1}\frac{1}{k}
\left (z^{-k}\frac{\p }{\p t_k}
+\bar z^{-k}\frac{\p }{\p \bar t_k}\right )
\eeq
Since the derivatives of $F_0$ with respect to the moments
$t_k$ are moments of the complimentary domain, this function
formally solves the 2D inverse potential problem.

\item
{\it 2-nd order derivatives:}
for $z, z' \in \Dc$ we have
\beq\label{2nd} {\cal D} (z) {\cal D} (z') F_0 =2G(z, z') -
\log \left | \frac{1}{z}-\frac{1}{z'}\right |^2
\eeq
where $G(z, z')$ is the Green function of the Dirichlet
boundary problem in $\Dc$ (\ref{Green}).
Note that the logarithmic singularity of the Green function
at $z=z'$ cancels by the second term in the right hand side.
In a particular case when both $z, z'$ tend to infinity,
we get a simple formula for the conformal radius:
\beq\label{cr}
\p_{t_0}^{2}F_0 =2\log r
\eeq

\item
{\it 3-d order derivatives. }
The generating formula reads \cite{MWZ}:
\beq\label{3d}
{\cal D} (a){\cal D} (b){\cal D} (c)F_0 =-\, \frac{1}{2\pi}
\oint_{\p \DD}
\p_n G(a, \xi ) \p_n G(b, \xi ) \p_n G(c, \xi )
|d \xi |
\eeq
An important corollary of this formula and
eq.\,(\ref{2nd}) is the complete symmetry of
the expression ${\cal D} (a)G(b,c)$ with respect to all
permutations of the points $a,b,c$.
Another corollary
of (\ref{3d}) is the following
{\it residue formula} valid for $j,k,l \geq 0$ \cite{BMRWZ}:
\beq\label{res}
\frac{\p^3 F_0}{\p t_{j} \p t_{k} \p t_{l}}=
\frac{1}{2\pi i}\oint _{|w|=1}
\frac{h_j (w) h_{k} (w) h_l (w) }{z'(w)\bar z' (w^{-1})}
\, \frac{dw}{w}
\eeq
Here $h_j (w)$
are polynomials in $w$ of degree $j$:
$$
h_j (w)=w\frac{d}{d w}
\left [ (z^j (w))_{+} \right ]
\;\; \mbox{for $j\geq 1$ and}
\;\;\;h_0 (w)=1\,,
$$
where $(...)_+$ is the positive degree
part of the Laurent series.
The notation $\bar z'(w^{-1})$ means the derivative
$d \bar z(u)/du$ taken at the point $u=w^{-1}$.
This formula is especially useful when $z'(w)$ is a rational
function, then the integral is reduced to a finite sum
of residues. We use this below.

\item
{\it Dispersionless Hirota equations.}
The function $F_0$ obeys an infinite number of
non-linear differential equations which are combined
into the integrable hierarchy of dispersionless Hirota's equations.
See \cite{KKMWZ,MWZ} for details.

\item
{\it WDVV equations.} Suppose $V(z)$ is a polynomial
of $m$-th degree, i.e., $t_k =0$ for all $k>m$.
On this subspace of parameters, $F_0$ obeys the system
of Wit\-ten-\-Dijkg\-raaf-\-Ver\-lin\-de-\-Ver\-lin\-de
(WDVV) equations
\beq\label{WDVV}
{\sf F}_i {\sf F}_{j}^{-1}{\sf F}_k=
{\sf F}_k {\sf F}_{j}^{-1}{\sf F}_i
\;\;\;\;
\mbox{for all $0\leq i,j,k \leq m-1$}
\eeq
where ${\sf F}_i$ is the $m$ by $m$ matrix
with matrix elements $({\sf F}_i )_{jk}=
\frac{\p^3 F_0}{\p t_i \p t_j \p t_k}$.
See \cite{BMRWZ} for details.

\end{itemize}

To conclude: $F_0$ is a ``master function''
which generates objects of
complex analysis
in planar simply-connected domains.
The full free energy of the matrix ensemble, $F$,
may be regarded as its ``quantization''.

\section{The genus 1 correction
to the free energy}

\subsection{The result for $F_1$}

We now describe the result  for the genus-1 correction
$F_1$. We start with
the special potential (\ref{Z2}). Then $F_1$ is expressed
entirely in terms
of the metric on the ${\sf U^c}$ induced from the
standard flat metric on the $z$-plane by the conformal map:
$dz \, d\bar z = e^{2\phi (w)}
dw \, d\bar w$. Here
\beq\label{f1}
\phi (w)=\log |z'(w)|
\eeq
and $z(w)$ is the conformal map ${\sf U^c} \to \Dc$ (\ref{zw}).
The derivation of this formula and its extension to a
general potential
is outlined in Section 5.

We found that
\beq\label{f2}
F_1 = -\, \frac{1}{24\pi}
\oint_{|w|=1} (\phi \p_n \phi +2 \phi ) \, |dw|
\eeq
Here $\p_n$ is the normal derivative, with the normal vector
pointing outside the unit circle.
The derivation of this formula is outlined, for a more general
model, in Section 5.

Since $\phi (w)$ is harmonic in ${\sf U^c}$, we may rewrite
the r.h.s. of (\ref{f2}) as
$$
F_1 =\frac{1}{24\pi}\int_{|w|>1}
|\nabla \phi |^2 d^2 w -
\frac{1}{12\pi}
\oint_{|w|=1} \phi  \, |dw|
$$
Here we recognize the
formula for the regularized
determinant of the Laplace operator
$\Delta_{\Dc}=4\p_z \p_{\bar z}$ in $\Dc$
with Dirichlet conditions on the boundary.
The first term is the bulk contribution first
found by Polyakov \cite{Polyakov} (for a metric
induced by a conformal map it reduces to a boundary integral),
while the second term,  computed
in \cite{Alvarez}, is a net boundary term
(see also Section 1 of \cite{OPS}):
\beq\label{f3}
F_1 =-\, \frac{1}{2} \log \det \left (- \Delta_{\Dc}\right )
\eeq
In the particular case $W(z)=-z\bar z$ we get
$F_1 =-\frac{1}{12}\log t_0$ that coincides with the
result of \cite{FGIL} obtained by a direct calculation.

The appearence of elements of quantum field theory in a
curved space is not accidental.
A field-theoretical derivation of this
result will be given elsewhere.

\subsection{Rational case}

Before explaining the origin of the explicit formula
for $F_1$ we write it in yet another suggestive form.
Consider a domain such that $z'(w)$ is a rational function:
$$
z'(w)=r\prod_{i=0}^{m-1}\frac{w-a_i}{w-b_i}
$$
All the points $a_i$ and $b_i$
must be inside the unit circle, otherwise the map
$z(w)$ is not conformal.
On the unit circle we have
$|dw| =\frac{dw}{iw}$
and $\phi (w) =\frac{1}{2}( \log z'(w) +
\log \bar z'(w^{-1}))$,
where the first and the second term
(the Schwarz reflection) are analytic
outside and inside it, respectively.
(Recall that our notation $\bar z'(w^{-1})$ means
$d \bar z(u)/du$ at the point $u=w^{-1}$.)
Plugging this into (\ref{f2}), we get:
$$
\begin{array}{ll}
F_1 \, =&\displaystyle{-\, \frac{1}{24\pi i}\oint_{|w|=1}
\log z'(w)
\left [ \frac{1}{2}\p_w \log z'(w)+\frac{1}{w}\right ] dw}\, -
\\&\\
&\displaystyle{- \, \frac{1}{24\pi i}\oint_{|w|=1}
\log \bar z'(w^{-1}) \frac{dw}{w} -
\frac{1}{48\pi i}\oint_{|w|=1}
\log \bar z'(w^{-1}) \frac{z''(w)}{z'(w)}dw}
\end{array}
$$
The integrals can be calculated by taking residues
either outside or inside the unit circle.
The poles are at $\infty$, at $0$, and at the points
$a_i$ and $b_i$.
The result is
\beq\label{PA1}
F_1 =-\frac{1}{24}\left ( \log r^4 +
\sum_{z'(a_i)=0}\log \bar z'(a_{i}^{-1})
-\sum_{z'(b_i)=\infty}\log \bar z'(b_{i}^{-1})
\right )
\eeq

If the potential $V(z)$ is polynomial,
$V(z)=\sum_{k=1}^mt_k z^k$,
i.e., $t_k =0$ as $k > m$ for some $m>0$, then
the series for the conformal map $z(w)$ truncates:
$z(w)= rw +\sum_{l=0}^{m-1}u_l w^{-l}$
and
$$
z'(w)=r\prod_{i=0}^{m-1}(1-a_iw^{-1})
$$
is a polynomial in $w^{-1}$
(all poles  $b_i$ of $z'(w)$
merge at the origin).
Then the
last sum in (\ref{PA1}) becomes
$m\log r$ and
the formula (\ref{PA1}) gives
\beq\label{eyn1}
F_{1}=-\frac{1}{24} \log \left (
r^4 \prod_{z'(a_j)=0}\frac{\bar z'(a_{j}^{-1})}{r}\right )=
-\frac{1}{24} \log \left (
r^4 \prod_{i,\,j=0}^{m-1}(1-\bar a_ia_j)\right )
\eeq
This formula is essentually identical  to the genus-1 correction
to the free energy of the Hermitian 2-matrix model
with a polynomial potential recently computed by
Eynard \cite{Eynard1}.

\subsection{Determinant representation of $F_1$
for polynomial potentials}

For polynomial potentials
the genus-1 correction enjoys an interesting
determinant representation.

Set
$$
D_m:=\det \left (\frac{\p^3 F_0}{\p t_{0} \p t_{j} \p t_{k}}
\right )_{0\leq j,k \leq m-1}
$$
Using the residue formula (\ref{res}) we compute:
\beq\label{det2}
D_m= \frac{1}{(2\pi i)^m}\oint_{|w_0|=1}
\frac{dw_0}{w_0} \, \ldots \, \oint_{|w_{m-1}|=1}
\frac{dw_{m-1}}{w_{m-1}}\,\,
\frac{\mbox{det}\,
\left [ h_j (w_j )h_k (w_j)\right ]}{\prod_{l=0}^{m-1}
z'(w_l)\bar z'(w_{l}^{-1})}
\eeq
Clearly, the determinant in the numerator can be substituted
by $\frac{1}{m}\det^2 (h_j(w_k))$ and
$\mbox{det}\,
\left [ h_j (w_k )\right ]=
(m-1)! \, r^{\frac{1}{2}m(m-1)} \Delta_m (w_i)$,
where $\Delta_m (w_i)$ is the Vandermonde determinant.
Each integral in (\ref{det2}) is given by the sum of residues
at the points $a_i$ inside the unit circle (the residues at
$w_i =0$ vanish).
Computing the residues and summing over all permutations
of the points $a_i$, we get:
\beq\label{det3}
D_m=(-1)^{\frac{1}{2}m(m-1)}((m-1)!)^2 \,
r^{m(m-3)}
\frac{\prod_{j}a_{j}^{m-1}}{\prod_{i,\,j}(1-a_j\bar a_{i})}
\eeq
As is seen from (\ref{tk}),
the last non-zero coefficient of $V(z)$ equals
$t_m = \frac{\bar u_{m-1}}{m r^{m-1}}$. (We regard it
as a fixed parameter.)
Therefore,
$\prod_{i=1}^{m}a_i = (-1)^m \, m(m\! -\! 1) r^{m-2}\bar t_m$,
and we represent $F_1$ (\ref{eyn1}) in the form
\beq\label{det4}
F_1 =\frac{1}{24}\log D_m
-\frac{1}{12}(m^2 \! -\! 3m\! +\! 3)\log r -
\frac{1}{24}(m\! -\! 1)\log \bar t_m +\mbox{const}
\eeq
where const is a numerical constant.
Recalling (\ref{cr}), we
see that $F_1$, for models with polynomial potentials
of degree $m$,
is expressed through derivatives of $F_0$:
\beq\label{det5}
F_1 =\frac{1}{24}\log\det_{m\times m}
\left (\frac{\p^3 F_0}{\p t_{0} \p t_{j} \p t_{k}}
\right )-\frac{1}{24}(m^2 \! -\! 3m\! +\! 3)
\, \frac{\p^2 F_0}{\p t_0^2}-
\frac{1}{24}(m\! -\! 1)\log \bar t_m +\mbox{const}
\eeq
where $j,k$ run from $0$ to $m-1$.

Similar determinant formulas are known
for genus-1 corrections to
free energy in topological field theories \cite{DZ}.

\section{$F_1$ from loop equation}

The standard (and powerful) method to obtain
$1/N$-expansions in matrix
models is to use  invariance of the
partition function under changes of matrix integration variables.
In the 2D Coulomb gas formalism,
this reduces to invariance of the partition function (\ref{Z1})
under diffeomorphisms
$$
z_i \longrightarrow z_i +\epsilon (z_i , \bar z_i)\,,
\;\;\;\;
\bar z_i \longrightarrow z_i +\bar \epsilon (z_i , \bar z_i)
$$
The  invariance of the partition function in
the first order in $\epsilon$
results in the identity
\beq\label{L0}
\sum_i
\int \p_{z_i}\left ( \epsilon (z_i , \bar z_i)\, e^{{\cal E}}\right )
\prod_j d^2 z_j =0
\eeq
for any function $\epsilon (z , \bar z)$.
One may read it as Ward
identitiy obeyed by correlation functions of the model.
For  historical reasons, it is called the loop equation.
Since correlation functions are variational derivatives
of the free energy with respect to the potential, the loop
equation is an implicit functional relation for the free
energy.

\subsection{Loop equation in general normal matrix model}

A closed loop equation does not emerge
for the special potential (\ref{Z2}).
It can be written only for the ensemble
of normal matrices with a general potential
$W$ in (\ref{Z1}).
Let it be of the form
$$
W(z)=-z\bar z +V(z)+\overline{V(z)} +U(z)
$$
where $U$ is only assumed to have
a regular Taylor expansion around the origin starting
from cubic terms.

Choosing $\epsilon (z_i , \bar z_i )=(z-z_i)^{-1}$ and
plugging it into (\ref{L0}) with ${\cal E}$ given in (\ref{energy}),
one is able to rewrite (\ref{L0}) as a relation between
correlation functions of the field
$$
\Phi (z)=-\hbar \, \mbox{tr} \, \log
\left [ (z-M)(\bar z -M^{\dag})\right ]
=-\hbar \sum_i \log |z-z_i |^2
$$
or $\p \Phi (z)=-\hbar \mbox{tr}(z-M)^{-1}$
(here and below $\p \equiv \p_z$).
Note that $\p \Phi (z)$
is trace of the resolvent of the matrix $M$ and
$\Delta \Phi (z)=-4\pi \rho (z)$, where $\rho$ is the
density of eigenvalues.
After some simple rearrangings, the loop equation
following from (\ref{L0}) acquires the form
\beq\label{loop1}
\frac{1}{2\pi}\int \frac{\p W(\zeta )
\left < \Delta
\Phi (\zeta )\right >}{z-\zeta}\, d^2 \zeta =
\Bigl < (\p \Phi (z) )^2 \Bigr >+
\hbar \Bigl < \p ^2 \Phi (z) \Bigr >
\eeq
(For any symmetric
function $f(\{z_i\})$, the correlation function
$\bigl <f \bigr >$ is defined as the
integral $\int f(\{z_i \})
|\Delta_N (z_i)|^2 \prod_j e^{\frac{1}{\hbar}W(z_j)}d^2 z_j$
with a normalization factor such that $\bigl < 1\bigr >=1$.)
This relation is exact for any finite $N$.
Supplemented by the relation
\beq\label{loop101}
\Bigl <\p \Phi (z)\Bigr >=-\frac{t_0}{z} +
\p_z {\cal D}(z)F
\eeq
(also exact) which directly follows from the definitions
of the free energy and the field $\Phi$, the
loop equation allows one to find the $\hbar$-expansion
of the free energy.

\subsection{Expanding the loop equation}

The $\hbar$-expansion of the free energy for
the general normal matrix model
is more complicated than the one discussed in the previous
sections.
It contains all powers of
$\hbar$, not only even:
\beq\label{L1}
\hbar^2 \log Z_N =F_0 +\hbar F_{1/2} +\hbar^2 F_1 + O(\hbar^3)
\eeq
so it hardly has a direct topological interpretation.
Accordingly, $\hbar$-expansions of mean values and other
correlation functions are expansions in $\hbar$ rather than
$\hbar^2$. In particular,
\beq\label{Phi1}
\left < \Phi (z)\right >=\Phi_0 (z)
+\hbar \Phi_{1/2} (z) +\hbar^2 \Phi_1 (z) +O(\hbar^3)
\eeq

We proceed by expanding the loop equation in powers of $\hbar$.
In the leading order, the second term
in the r.h.s. vanishes, and $\bar z$-derivative of
the both sides gives:
\beq\label{exp2}
(\p W(z) -\p \Phi_0 (z))\, \Delta \Phi_0 (z) =0
\eeq
This just means that for $z\in \DD$ the
extremum condition (\ref{semitau1})
is satisfied
and $\Delta \Phi_0 (z) =0$ otherwise.
Inside $\DD$, the leading term of the
mean density, $\rho_0 (z)$,
is given by
$\rho_0 (z)=\sigma (z)$, where $\sigma (z)$
is defined in (\ref{sigma}).
(Note that the function $\sigma$ is defined by this formula
everywhere in the complex plane, and does not depend on the
shape of $\DD$, while $\rho_0$ coincides with $\sigma$
in $\DD$ and equals $0$ in $\Dc$.) For potentials
of the form (\ref{Z2}), $\sigma (z)=1/\pi$.

Being developed into
a series in $\hbar$, the loop equation
gives an iterative procedure to find the
coefficients $\Phi_i (z)$.
We need the following results on the correlation
functions for the general normal matrix ensemble
(see \cite{WZ2}):
\beq\label{result1}
\Bigl <\p \Phi (z)\Bigr >=
\int_{\DD}\frac{\sigma (\zeta )d^2 \zeta}{\zeta -z}
\, +\, O(\hbar )
\eeq
\beq\label{result2}
\Bigl <\Phi (z_1) \Phi (z_2) \Bigr >_{{\rm conn}}=
2\hbar^2 \left (
G(z_1 , z_2) - G(z_1 , \infty ) - G(\infty , z_2 )
-\log \frac{|z_1 -z_2 |}{r}\right ) +O(\hbar^3)
\eeq
where the connected
correlation function is defined as
$\bigl < fg\bigr >_{{\rm conn}}=
\bigl < fg\bigr >-
\bigl <f\bigr > \bigl <g\bigr >$.
Note that the function (\ref{result2}) has
no singularity at coinciding points
$z_1 , z_2 \in \Dc$. Merging the points,
we get:
\beq\label{result3}
\Bigl <(\p \Phi (z))^2\Bigr >_{{\rm conn}}=
\frac{\hbar^2}{6}\,
\{ w ; \, z\} +O(\hbar^3)
\eeq
where
$$
\{ w ; \, z\} =\frac{w'''(z)}{w'(z)}
-\frac{3}{2}\left (\frac{w''(z)}{w'(z)}\right )^2
$$
is the Schwarzian derivative of the conformal map
$w(z)$.

After these preparations,
further steps are straightforward. Terms of order
$\hbar$ and $\hbar^2$ of the loop equation give:
\beq\label{op3}
\begin{array}{l}
\displaystyle{
\frac{1}{2\pi}\int L(z, \zeta )\, \Delta \Phi_{1/2}(\zeta )
\,d^2 \zeta =
-\p^2 \Phi_0 (z)}
\\ \\
\displaystyle{
\frac{1}{2\pi}\int L(z, \zeta )\, \Delta \Phi_{1}(\zeta )
\,d^2 \zeta =
-\left [ \Bigl (\p \Phi_{1/2} (z) \Bigr )^2
+\p^2 \Phi_{1/2}(z)\right ]-\frac{1}{6}\{w; \, z\} }
\end{array}
\eeq
where the kernel of the integral operator in the l.h.s. is
\beq\label{kernel}
L(z, \zeta )=
\frac{\p W(\zeta )-\p \Phi_0 (z)}{\zeta -z}
\eeq

It should be noted that the $\hbar$-expansion
of the loop equation
may break down for $z\in \DD$.
This is mainly because
the correlator $\bigl <\Phi (z)\Phi (z')\bigr >_{{\rm}}$,
when the two points
are close to each other and
belong to the support of eigenvalues,
is not given by eq.\,(\ref{result2}).
At the same time, for our purpose we need
this correlator just
on very small distances, when the two points merge.
Naively, for $z,z'\in \DD$
the correlator diverges as $z'\to z$.
This means that
its short-distance behaviour
is in fact of a different order in $\hbar$ and
must be calculated separately. Fortunately, this problem
can be avoided by restricting the equations
to $\Dc$, where no
divergency emerges on any scale
and one may think that the short-distance
behaviour of correlation functions is still given
by eq.\,(\ref{result2}). (However, we understand
that this argument is not rigorous and need to be
justified by an actual calculation of correlation
functions at small scales.) Hereafter, $z$ in (\ref{op3})
is assumed to be outside the support of eigenvalues,
i.e., the equations should be solved for $z\in \Dc$.
In this region the functions $\Phi_k (z)$ are harmonic.

From (\ref{loop101})
we see that
\beq\label{deriv}
\p_z {\cal D}(z)F_{1/2}=\p_z \Phi_{1/2}(z)\,,
\;\;\;\;\;
\p_z {\cal D}(z)F_{1}=\p_z \Phi_{1}(z)
\eeq
The strategy is to find $\Phi_k$'s from (\ref{op3}) and
then ``to integrate'' them to get $F_k$'s, i.e.,
to find a functional $F_k$
such that it obeys (\ref{deriv}). A general method to find
the ``derivative''
${\cal D}(z)$ of any proper functional of the domain $\Dc$
is proposed in \cite{MWZ}.

An important remark is in order.
Suppose we restrict ourselves to
the class of models with potentials
of the form (\ref{Z2}) (i.e.,
with $\sigma (z)=1/\pi$), like in previous sections.
Applying $\p_z {\cal D}(z)$ to the functional
(\ref{f2}), that is $F_1$ in this case,
we obtain a wrong
answer for $\p_z \Phi_1 (z)$, which
does not obey the loop equation
(\ref{op3})! This seemingly contradicts
eqs.\,(\ref{deriv}) and so explains why one has to
deal with the arbitrary
potential. The matter is simply that
there are functionals such that they
vanish for potentials with $\sigma (z)=1/\pi$ but
their ``derivatives'', $\p_z {\cal D}(z)$, do not.
They do contribute to $\Phi_1$ and restore the right
answer.

\subsection{Free energy of the general model}

Skipping further details, we present the results
for the general model of normal matrices.

The answer for $F_0$ is familiar \cite{WZ2}.
It is given by (\ref{F0}).
The first correction, $F_{1/2}$, is
\beq\label{ext2}
F_{1/2} =-\,
\int_{\DD}
\sigma (z) \log \sqrt{\pi \sigma (z )}
\, d^2 z
\eeq

To write down the full answer for $F_1$ in a compact form,
we need to introduce, along with the
$\phi (w)$ (\ref{f1}), another function,
\beq\label{chi}
\chi (z)=\log \,  \sqrt{\pi \sigma (z)}
\eeq
and the function
$\chi^H (z)$ defined in the domain $ \Dc$. It is a
harmonic function in $\Dc$ with the boundary value
$\chi (z)$. The function
$\chi^H$ is the
solution to the Dirichlet boundary problem:
$\chi^H (z)=-\frac{1}{2\pi}\, \int_{\p \DD}
\p_n G(z, \xi)\chi (\xi) |d\xi |$.
The explicit formula for $F_1$ reads:
\beq\label{ff3}
\begin{array}{lll}
F_1 &=& \displaystyle{
\frac{1}{24\pi} \left [
\int _{|w|>1} |\nabla (\phi +\chi ) |^2 \, d^2 w
-2\oint _{|w|=1} \! (\phi +\chi ) \, |dw|
-\int_{{\bf C}} |\nabla \chi |^2 \, d^2 w
\right ]} \, +
\\ &&\\
&&\displaystyle{
+\, \frac{1}{8\pi}\left [ \int _{\DD} |\nabla \chi |^2
d^2 z -\oint _{\p \DD}\chi \p_n \chi^H \, |dz|
-\frac{1}{2}\int _{\DD} \Delta \chi d^2 z \right ]}
\end{array}
\eeq
where $\chi$ in the first three integrals is treated
as a function of $w$ through $\chi = \chi (z(w))$.

The r.h.s. of this formula is naturally decomposed into
two parts having completely different nature,
the ``quantum'' and ``classical'' parts:
$F_1 =F_{1}^{({\rm q})}+F_{1}^{({\rm cl})}$.
The (most interesting) quantum part can be again represented
through the regularized
determinant of the Laplace operator in $\Dc$ with
Dirichlet boundary conditions. However,
now the Laplacian should be taken in conformal metric
with the conformal factor $e^{2\chi (z)}$.
Equivalently, on the exterior
of the unit circle the Laplacian
should be taken in the metric with the conformal factor
$e^{2\phi (w)+2\chi (z(w))}$; we see that $\phi$ and $\chi$
do enter as the sum $\phi +\chi$ in the first line.
More precisely, the formula for regularized determinants
of Laplace operators in domains with boundary known in
the literature
(eq.\,(4.42) in \cite{Alvarez})
allows us to identify
\beq\label{F1q}
F_{1}^{({\rm q})}=
\frac{1}{2}\log \, \frac{\det
\left (-e^{-2\chi}\Delta_{{\bf C}}\right )}{\det \left (-e^{-2\chi}
\Delta_{\Dc}\right )}
\eeq
The classical part comes from ``classical''
(though of order $\hbar$) corrections to the shape
of the support of eigenvalues, which always
exist unless $\sigma (z)$ is a constant (see below).
It is essentially given
by $F_{1}^{({\rm cl})}=\lim_{\hbar \to 0}
\bigl ( \frac{1}{2\hbar^2}\bigl <
(\mbox{tr}\, \chi (M))^2 \bigr >_{{\rm conn}}\bigr )$.

Different terms of the $\hbar$-expansion gain
a clear interpretation in terms of
the collective field theory of the normal matrix model,
in the spirit
of \cite{Jevicki}.
In this context, it is natural to
start with the general Coulomb
gas model with arbitrary $\beta$.
The generalized loop equation
\beq\label{col2}
\frac{1}{2\pi}\int \frac{\p W(\zeta )
\left < \Delta
\Phi (\zeta )\right >}{z-\zeta}\, d^2 \zeta =
\beta \Bigl < (\p \Phi (z) )^2 \Bigr >+
(2-\beta )\hbar \Bigl < \p ^2 \Phi (z) \Bigr >
\eeq
can be understood
as the conformal Ward identity
for the collective theory.
This allows one
to find the effective action in the form
\beq\label{col3}
\begin{array}{l}
S=S_0 +S_1
\\ \\
\displaystyle{
S_0 =\beta \int \!\! \int \rho (z)
\log |z-\zeta |\rho (\zeta )\, d^2 z d^2 \zeta +
\int W(z)\rho (z)\, d^2 z}
\\ \\
\displaystyle{
S_1 =\Bigl (1-\frac{\beta}{2}\Bigr )
\hbar \int \rho (z)\log \rho (z)\, d^2 z}
\end{array}
\eeq
The second term, $S_2$,
is a combination of the short range part
$-\frac{\beta}{2}\rho \log \rho$ and the entropy $\rho \log \rho$.
(See \cite{Dyson,herm-col}, where similar actions
for unitary and Hermitian matrix ensembles were discussed.)

This action suggests to rearrange
the $\hbar$-expansion of the free energy (\ref{L1}) and
write it in the ``topological'' form
$F=\sum_{g\geq 0}\hbar^{2g}F_g$, where each term
has its own expansion
\beq\label{col4}
F_g =F_{g}^{(0)}+\sum_{n\geq 1}
\hbar_{\beta}^{n}F_{g}^{(n)}
\eeq
Here $\hbar_{\beta}\equiv (2-\beta) \hbar$
is regarded as an independent parameter.
The equilibrium density of charges, $\rho_0$, is determined
by $\delta S/\delta \rho =0$ which leads to
the Liouville-like equation
\beq\label{col5}
-\frac{\hbar_{\beta}}{8\pi} \Delta \log \rho_0 (z)+
\beta \rho_0 (z)=\sigma (z)
\eeq
in the bulk. For $\beta \neq 2$ the first term generates corrections to the
shape of the support of eigenvalues.
The classical free energy is
$F_0=F_{0}^{(0)}+\hbar_{\beta}
F_{0}^{(1)}+\hbar_{\beta}^{2} F_{0}^{(2)}+O(\hbar_{\beta}^{3})$.
In particular we see that $F_{1/2}$ given in (\ref{ext2})
is in fact $F_{0}^{(1)}$ while the ``classical'' part
$F_{1}^{({\rm cl})}$ of (\ref{ff3}) is
$F_{0}^{(2)}$. The ``quantum'' part is then
$F_{1}^{({\rm q})}=F_{1}^{(0)}$.

The collective field theory approach to the normal
and complex matrix ensembles will be presented elsewhere.

\section*{Acknowledgments}

We thank A.Cappelli, L.Che\-khov, B.Dub\-ro\-vin, V.Ka\-za\-kov,
I.Kos\-tov,
Yu.Ma\-ke\-en\-ko, A.Mar\-sha\-kov, M.Mi\-ne\-ev-\-Wein\-stein
and R.Theo\-do\-res\-cu
for useful discussions.
P.W. would like to thank INFN and the
University of Florence, Italy, for hospitality.
A.Z. is grateful to B.Julia
for opportunity to present these results
at the Les Houches Spring School in March 2003.
P. W. was supported by the NSF MRSEC Program under
DMR-0213745, NSF DMR-0220198 and Humboldt foundation.
The work of A.Z. was supported in part by RFBR grant
03-02-17373, by grant for support of scientific schools
NSh-1999.2003.2 and by Federal Program of the Russian
ministry of industry, science and technology 40.052.1.1.1112.
The work of both authors was partially supported by the
NATO grant PST.CLG.978817.

\end{document}